# ROLE OF ΛN AND ΛNN INTERACTION PARAMETERS ON BINDING ENERGY OF $^4_\Lambda H$ AND $^4_\Lambda H^*$

Bhupali Sharma*
*Dept. of Physics, Arya Vidyapeeth College, Guwahati-781016*
*For correspondence. (bhupalisharma@gmail.com)

Abstract: Variational Monte Carlo study has been done for the two hypernuclear systems $^4_\Lambda H$ and $^4_\Lambda H^*$ for calculation of binding energies. For the two hypernuclear systems under study, different potential models have been used for the interactions involved in these hypernuclear systems. ArgonneV$_{18}$ NN, Urbana IX NNN and phenomenological ΛN and ΛNN potentials have been used in our study. Our potential models are based on our previous studies on different double lambda hypernuclear systems. From our results, hyperon-nucleon parameters ie. two-body ΛN parameter and three-body ΛNN parameters are found to be important for binding the hypernuclear systems under study. With reduction in the values of ΛNN interaction parameters used in earlier works, there is significant difference in the values of binding energy of $^4_\Lambda H$ and $^4_\Lambda H^*$. Also, ΛN interaction parameters are also found to play important role in binding.

1. Introduction:

A hypernucleus is a nucleus which contains one or more hyperon in addition to the nucleons. The first report of hypernuclear event was made by M. Danysz and J. Pniewski[1] in 1953. Since then many confirmed single and double hypernuclear events have been reported in various experiments. To study these confirmed as well as undiscovered hypernuclear systems, many theoretical studies have also been done on different single and double hypernuclear systems since the first report.

We, have done Variational Monte Carlo studies in our previous works on different hypernuclei using our preferred potentials and found the influence of ΛN and ΛNN potential parameters to be crucial for binding double lambda hypernuclear system[2,3,4]. In these studies we used three preferred potential models viz. ΛN1, ΛN2 and ΛN3 with different ΛN and ΛNN potential parameters. We also found that, the binding energy of different hypernuclear systems for our preferred potential models depend crucially on three-body ΛNN parameters and on the exchange part of ΛN interaction.

In the present study we report few more results on the hypernuclear systems $^4_\Lambda H$ and $^4_\Lambda H^*$ using two new potential model with ΛN and ΛNN potential parameters different from our previous studies. We call these ΛN and ΛNN potential models as ΛN4 and ΛN5.

Earlier many works have been done on $^4_\Lambda H$ and $^4_\Lambda H^*$. New experimental results for $^4_\Lambda H$ with binding energy value of 2.12 MeV have been reported recently in the first high-resolution pion spectroscopy from decays of strange systems done at Mainz Microtron MAMI[5]. Recently, more theoretical studies also have been done on these two hypernuclear systems [6,7,8].

2. Hamiltonian and wavefunction:

We use *ArgonneV$_{18}$* NN[9] and *Urbana IX* NNN[10] potentials for the nuclear part of the Hamiltonian [2]. For ΛN potential, we use phenomenological potential consisting of central, Majorana space-exchange and spin-spin ΛN components and is given by,

$$V_{\Lambda N} = (V_c(r) - \overline{V}T_\pi^2(r))(1 - \varepsilon + \varepsilon P_x) + \frac{1}{4}V_\sigma T_\pi^2(r)\sigma_\Lambda \cdot \sigma_N \tag{1}$$

where $P_x$ is the majorana space-exchange operator and $\varepsilon$ is the space exchange parameter which is taken as 0.2[11]. $V_c(r)$, $\overline{V}$ and $V_\sigma$ are respectively Wood-saxon core, spin-average and spin-dependent strength and $T_\pi^2(r)$ is one-pion tensor shape factor.

The $\Lambda$NN potential consists of two terms. Firstly, a two-pion exchange and a dispersive part[12]. The two-pion exchange part of the interaction is given by

$$W_p = -\frac{1}{6} C_p (\tau_i . \tau_j)\{X_{i\Lambda} . X_{j\Lambda}\} Y_\pi(r_{i\Lambda}) Y_\pi(r_{j\Lambda}) \quad (2)$$

Where $X_{k\Lambda}$ is the one-pion exchange operator given by,

$$X_{k\Lambda} = (\sigma_k . \sigma_\Lambda) + S_{k\Lambda}(r_{k\Lambda}) T_\pi(r_{k\Lambda})$$

with

$$S_{k\Lambda} = \frac{3(\sigma_k . r_{k\Lambda})(\sigma_\Lambda . r_{k\Lambda})}{r_{k\Lambda}^2} - (\sigma_k . \sigma_\Lambda)$$

The dispersive part of the $\Lambda$NN potential is given by,

$$V_{\Lambda NN}^{DS} = W_0 T_\pi^2(r_{i\Lambda}) T_\pi^2(r_{j\Lambda})[1 + \frac{1}{6}\sigma_\Lambda . (\sigma_i . \sigma_j)] \quad (3)$$

$Y_\pi(r_{k\Lambda})$ and $T_\pi(r_{k\Lambda})$ are the usual Yukawa and tensor functions with pion mass, $\mu$=0.7 fm$^{-1}$. $C_p$ and $W_0$ are $\Lambda$NN interaction parameters.

The $\Lambda$N and $\Lambda$NN potential parameters for our preferred models[2] are listed in Table1. $C_p$ and $W_0$ are the strength parameters of the two-pion and dispersive parts of the $\Lambda$NN potential.

Table 1: $\Lambda$N and $\Lambda$NN interaction parameters. Except for $\varepsilon$, all other quantities are in MeV.

| $\Lambda N$ | $\overline{V}$ | $V_\sigma$ | $\varepsilon$ | $C_p$ | $W_0$ |
|---|---|---|---|---|---|
| $\Lambda N1$ | 6.150 | 0.176 | 0.2 | 1.50 | 0.028 |
| $\Lambda N2$ | 6.110 | 0.000 | 0.0 | 1.50 | 0.028 |
| $\Lambda N3$ | 6.025 | 0.000 | 0.0 | 0.00 | 0.000 |

In this present work we do calculations on the selected hypernuclei using two new potential models with different $\Lambda$N and $\Lambda$NN interaction parameters, viz. $\varepsilon$, $C_p$ & $W_0$. These values of $C_p$ & $W_0$ were selected on the basis of giving bound state for $^3_\Lambda H$. The potential models used are listed in Table 2. For the two potential models, the spin-average and spin-dependent strength of the $\Lambda$N potential are kept same with spin-average strength $\overline{V}$ = 6.150 Mev and spin-dependent strength $V_\sigma$=0.176 Mev, same as in $\Lambda$N1[2,3].

Table 2: New $\Lambda$N and $\Lambda$NN interaction parameters. Except for $\varepsilon$, all other quantities are in Mev.

| $\Lambda N$ | $\overline{V}$ | $V_\sigma$ | $\varepsilon$ | $C_p$ | $W_0$ |
|---|---|---|---|---|---|
| $\Lambda N4$ | 6.150 | 0.176 | 0.2 | 0.70 | 0.012 |
| $\Lambda N5$ | 6.150 | 0.176 | 0.0 | 0.00 | 0.000 |

The variational wave function is of the form,

$$|\Psi_v\rangle = \left[1 + \sum_{i<j<k}(U_{ijk} + U_{ijk}^{TNI}) + \sum_{i<j,\Lambda} U_{ij,\Lambda} + \sum_{i<j} U_{ij}^{LS}\right] \prod_{i<j<k} f_c^{ijk} |\Psi_p\rangle \quad (4)$$

where, $|\Psi_p\rangle$ is the pair wave function[2,3] given by

$$|\Psi_p\rangle = S \prod_{i<j}(1+U_{ij}) S \prod_{i<\Lambda}(1+U_{i\Lambda}) |\Psi_J\rangle \quad (5)$$

The Jastrow wave function for lambda hypernuclei is given by,

$$|\Psi_J\rangle = [\prod_{i<j<k} f_c^{ijk} \prod_{i<\Lambda} f_c^{i\Lambda} \prod_{i<j} f_c^{ij}]|\Psi_{JT}\rangle |\varphi\rangle \quad (6)$$

Journal of Applied and Fundamental Sciences

where f's are the central correlation functions and $|\varphi\rangle$ is an antisymmetric wave function of the lambda particle. $|\Psi_\pi\rangle$ is the spin and isospin wavefuntion of the s-shell nucleus.

3. Technique:

Variational Monte Carlo method is used to find the ground state energy and binding energy of different hypernuclear systems. A suitably parametrized trial wave function is selected which is a function of position, spin, isospin and other intrinsic variables and parameters and this trial wave function is used to find the upper bound to the energy using Metropolis algorithm[13]. In this process, an initial random walk is made with the trial wave function to generate a set of configurations which are stored. Energy expectation values are calculated using the trial wave function, varying variational parameters one or two at a time. The energy expectation values are sampled both in configuration space and in the order of operators in the wave function by following a Metropolis random walk. The wavefunction that gives the lowest energy is then selected by Metropolis algorithm and is used to generate new configurations and the process is repeated till the lowest possible value of energy is found. The minimum energy is searched by calculating energy difference, ΔE (which is the difference in energies with old configurations and new configurations), for wave functions using configurations generated by random walk. If ΔE<0, the new configurations are accepted and the search for lowest possible value of energy is continued. The lowest value of energy calculated in this way is taken as true ground state energy in accordance with variational principle.

The variational principle states that the approximate value of a Hamiltonian, calculated using trial wave-function is never lower in value than the true ground state energy

$$E = \frac{\langle \Psi|H|\Psi \rangle}{\langle \Psi| \Psi \rangle} \geq E_o \tag{7}$$

The binding energy $B_\Lambda$ of a single hypernuclear system is given by,

$$-B_\Lambda(^A_\Lambda Z) = E(^A_\Lambda Z) - E(^{A-1}Z) \tag{8}$$

4. Results and discussion:

The binding energy results for the hypernuclear systems $^4_\Lambda H$ and $^4_\Lambda H^*$ with the potential models ΛN4 and ΛN5 are tabulated in Table 3. We have also presented the results for $^3_\Lambda H$ the two potentials.

Table3: Binding energy($B_\Lambda$) Results for $^4_\Lambda H$ and $^4_\Lambda H^*$ for different ΛN interactions. All quantities are in MeV.

| Potential | $^3_\Lambda H$ | $^4_\Lambda H$ | $^4_\Lambda H^*$ |
|---|---|---|---|
| *ΛN4* | **0.17(01)** | **1.89(02)** | **1.11(04)** |
| *ΛN5* | **0.15(00)** | **2.39(01)** | **1.83(01)** |
| *Experimental* | 0.13 | 2.12 | 1.12 |
| *ΛN1* | 0.34(01)[3] | 2.15(02)[2] | 1.06(02)[2] |

The potential model ΛN4 contains both space exchange part of ΛN potential and non zero values of the parameters $C_p$ & $W_0$ of ΛNN potential . For both ΛN4 and ΛN5, the binding energy for $^3_\Lambda H$ is more close to the experimental value compared to our earlier potential model ΛN1 [2,3,4].This is because the parameters in the potential model ΛN1 was fitted to the experimental value of the double hypernucleus $^6_{\Lambda\Lambda}He$[14]. In Table 4, we present the detailed results including space exchange contribution (SEC) and energy due to ΛN and ΛNN potentials for the potential model ΛN4 as it contains non zero values of ΛNN interaction parameters in addition to the space exchange parameter of ΛN potential.

Table 4: Detailed result for potential model ΛN4 for the hypernuclear systems $_\Lambda^4 H$ and $_\Lambda^4 H^*$. All quantities are in Mev.

|  | $_\Lambda^4 H$ | $_\Lambda^4 H^*$ |
|---|---|---|
| E | -10.21(02) | -9.43(04) |
| SEC | 0.19(01) | 0.20(01) |
| $V_{\Lambda N}$ | -10.61(16) | -9.94(29) |
| $V_{\Lambda NN}$ | -0.13(01) | 0.02(01) |
| $B_\Lambda$ | 1.89(02) | 1.11(04) |

5. Conclusions:

With potential model ΛN4, which has reduced values of ΛNN interaction parameters compared to ΛN1, the value of binding energy for $_\Lambda^4 H$ differs from the experimental value whereas for $_\Lambda^4 H^*$, the binding energy value agrees well with the experimental value(Table 3). Therefore reduction in the values ΛNN interaction parameters affects $_\Lambda^4 H$ but not $_\Lambda^4 H^*$. With ΛN5, for which exchange part of ΛN potential ε and ΛNN interaction parameters $C_p$ & $W_0$ are absent, the binding energy for both the hypernuclear systems are found to differ from experimental value. This is similar to earlier results with potential model ΛN2 and ΛN3[4].

Acknowledgements:

Bhupali Sharma acknowledges the funding offered by UGC for carrying out the work.

References:

[1] M. Danysz and J. Pniewski, Phil. Mag. 44, 348 ,1953.
[2] Q.N. Usmani, A.R. Bodmer and Bhupali Sharma, Phys. Rev. C **70,** 061001(R),2004.
[3] Bhupali Sharma, Chin. Phys. Lett., **30, 3** , 032101,2013.
[4] Bhupali Sharma, Proceedings of Xth Biennial National Conference of PANE, 51 ,2017.
[5] A. Esser et.al.Phys. Rev. Lett. 114, 232501,2015.
[6] A Nogga, Nucl. Phys. A, 914, 140-150 ,2013.
[7] M. Imran et. al., Journal of Phys. G, 41, 6 ,2014.
[8] D. Gazda and A. Gal, Nucl. Phys. A 954 , 161-175,2016.
[9] R.B. Wiringa et. al.,Phys. Rev C 51,1995.
[10] B.S. Pudliner et. al.,Phys. Rev Lett. 74,1995
[11] I.N. Filikhin and A. Gal, Phys. Rev. C **65**,041001,2002.
[12] Rita Sinha, Q.N. Usmani and B.M. Taib, Phys. Rev C **66**, 024006,2002.
[13] N. Metropolis et. al., J. Chem. Phys. 21, 6, 1953.
[14]H Takahashi et. al., Phys. Rev. Lett. 87, 212502, 2001.